\documentclass[%
reprint,
superscriptaddress,
 amsmath,amssymb,
 aps,
]{revtex4-1}

\usepackage{graphicx}
\usepackage{dcolumn}
\usepackage{bm}
\usepackage{amsmath}
\usepackage[utf8]{inputenc}
\usepackage[T1]{fontenc}
\usepackage{mathptmx}
\usepackage{textcomp}
\usepackage{gensymb}
\usepackage{tensor}
\usepackage{xcolor}
\usepackage{braket}
\usepackage[normalem]{ulem}

\preprint{APS/123-QED}
\begin{document}
\title{Two-tone Doppler cooling of radial two-dimensional crystals in a radiofrequency ion trap}
%
\affiliation{University of Washington, Department of Physics, Seattle, Washington, USA, 98195}
\affiliation{St. Olaf College, Northfield, Minnesota, USA, 55057}
\affiliation{Physics Department, Brookhaven National Laboratory, Upton, New York, USA, 11973}

\author{Alexander Kato}
\affiliation{University of Washington, Department of Physics, Seattle, Washington, USA, 98195}
\author{Apurva Goel}
\affiliation{University of Washington, Department of Physics, Seattle, Washington, USA, 98195}
\author{Raymond Lee}
\affiliation{University of Washington, Department of Physics, Seattle, Washington, USA, 98195}
\author{Zeyu Ye}
\affiliation{University of Washington, Department of Physics, Seattle, Washington, USA, 98195}
\author{Samip Karki}
\affiliation{St. Olaf College, Department of Physics, Northfield, Minnesota, USA, 55057}
\author{Jian Jun Liu}
\affiliation{Physics Department, Brookhaven National Laboratory, Upton, New York, USA, 11973}
\author{Andrei Nomerotski}
\affiliation{Physics Department, Brookhaven National Laboratory, Upton, New York, USA, 11973}

\author{Boris B. Blinov}
\affiliation{University of Washington, Department of Physics, Seattle, Washington, USA, 98195}
\date{\today}

\begin{abstract}
We study the Doppler-cooling of radial two-dimensional (2D) Coulomb crystals of trapped barium ions in a radiofrequency trap. Ions in radial 2D crystals experience micromotion of an amplitude that increases linearly with the distance from the trap center, leading to a position-dependent frequency modulation of laser light in each ion's rest frame.  We use two tones of Doppler-cooling laser light separated by approximately 100~MHz to efficiently cool distinct regions in the crystals with differing amplitudes of micromotion. This technique allows us to trap and cool more than 50 ions populating 4 shells in a radial two-dimensional crystal, where with a single tone of Doppler cooling light we are limited to 30 ions in 3 shells. We also individually characterize the micromotion of all ions within the crystals, and use this information to locate the center of the trap and to determine the Matthieu parameters $q_{x}$ and $q_{y}$. 

\end{abstract}

\maketitle
\section{Introduction}
Two-dimensional crystals of trapped ions represent a natural way to scale up the capabilities of trapped ions in a one-dimensional (1D) chain~\cite{Cirac1995,Monroe1995}. Having more nearest neighbours may enable a higher error threshold for fault-tolerant quantum computing~\cite{Porras2006,Wang2015}. In addition, quantum simulation of 2D systems such as spin liquids~\cite{Balents2010}, frustrated systems~\cite{Moessner2000}, quantum magnetism~\cite{Nath2015}, and spin-spin interaction~\cite{Welzel2011,Espinoza2021} may be more well suited to a native 2D geometry.

\begin{figure}[h]
\includegraphics[width=90mm]{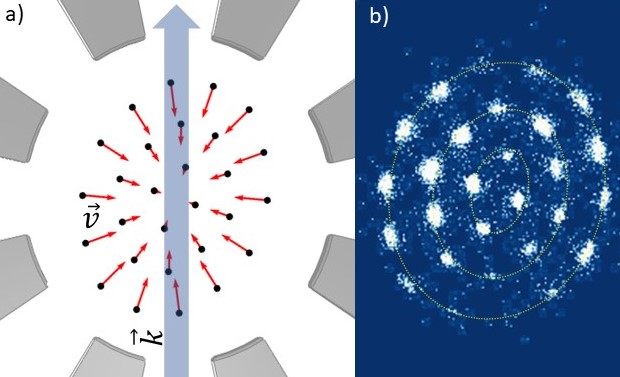}
\caption{\label{fig:Figure_1} Two-dimensional radial ion crystal in a sectored-ring ion trap. a) Schematic of laser cooling of a 28-ion 2D crystal undergoing "breathing" oscillations due to the micromotion. The ions' positions are indicated by the black dots, while the instantaneous micromotion velocity vectors $\vec{v}$ are shown with the arrows. The large transparent light-blue (grey) arrow indicates  $\vec{k}$ for the Doppler-cooling laser beam. b) EMCCD image of the 28 ion crystal depicted in (a). Doppler shifts $\vec{k}\cdot\vec{v}$ lead to uneven fluorescence profile over the ion crystal. Note that the ions in 2D Coulomb crystals form "shells", with three shells indicated in this particular crystal by the dashed yellow (white) lines.}
\end{figure}

2D crystals of trapped ions can be created in both radiofrequency (RF) traps and Penning traps. In Penning traps, 2D crystals are actively being pursued as a platform for quantum simulations and quantum information processing, with demonstrations of entanglement between hundreds of ions in a 2D lattice~\cite{Britton2012,Sawyer2012,Gilmore2021}, and near-ground state cooling in a 100-ion crystal~\cite{Jordan2019}. However, ions in Penning traps constantly rotate at  frequencies of order 100's of kHz, which makes individual qubit addressing difficult. In RF traps, ions in a 2D crystal are nearly stationary, which is desirable for scalable individual qubit operations. Yet, ions oscillate about their equilibrium positions at the RF drive frequency $\Omega$, called the micromotion. 

Proposed schemes to use 2D crystals of trapped ions in RF traps for quantum information applications exploit symmetry in the transverse direction, where micromotion can be compensated and minimized for every ion in the crystal~\cite{Yoshimura2015}. Lasers addressing ions in this direction are not Doppler-shifted in the ion's rest frame, and high fidelity one- and two-qubit operations can theoretically be achieved~\cite{Wang2015}. However, these operations require that the crystal can be effectively cooled and stabilized. The presence of large, position-dependent micromotion in the plane of the crystal makes the regular monochromatic Doppler cooling inefficient for large 2D crystals. This concept is illustrated in Figure 1.  A 28-ion crystal with individual instantaneous velocities ($\vec{v}$) due to the micromotion are shown in Figure 1(a), along with the direction of the Doppler-cooling laser. Ions that have small velocity component along the laser k-vector ($\vec{k}$) interact more efficiently with the laser and scatter more photons, which leads to better cooling of those ions. This can be seen in Figure 1(b), where the ions that scatter the Doppler-cooling light more efficiently appear as brighter spots.

2D crystals in RF traps can be produced in radial~\cite{Ivory2020,D'Onofrio2021} or lateral geometries~\cite{Wang2020,Yan2016}. In the radial geometry, the RF electric field (E-field) has cylindrical symmetry. 2D crystals can be formed in the plane where the transverse (E-field) has a node, and micromotion in the transverse direction can be minimized for all ions in the crystal. Radial 2D crystals were first explored over 30 years ago for up to 15 ions~\cite{Itano1988}, and more recently up to $\sim30$ ions~\cite{Ivory2020,Xie2021}. The planar and transverse normal modes have been shown to be well decoupled, and low heating rates have been demonstrated~\cite{D'Onofrio2021}.

Lateral 2D crystals have been studied since the introduction of the linear Paul trap \cite{Raizen1992}, leading to investigations of normal mode structure~\cite{Kaufmann2012} and demonstrations of sub-Doppler cooling techniques~\cite{Joshi2020}. Recent improvements in microfabricated ion trap capabilities have led to the formation of 2D crystals that can be cooled near the motional ground state~\cite{Qiao2021}. An advantage of lateral 2D crystals is the ability to cool ions along the trap axis where micromotion is largely non-existent, and larger 2D crystals have been reported using this method~\cite{Szymanski2012}. Both types of geometries continue to be studied as the potential platforms for scaling up 2D ion crystals for quantum information applications.

We study radial 2D crystals, where the influence of micromotion on Doppler cooling cannot be neglected. Micromotion causes a time-dependent Doppler shift of cooling lasers in the ion's rest frame.  For a 3-level atom such as Ba$^{+}$ (see inset in Figure 2), this causes the laser detunings $\Delta_{1}$ ($\Delta_{2}$) of the 493~nm (650~nm) to vary according to the instantaneous velocity of the ion. Consider an ion oscillating in an RF trap at $\Omega=2\pi\times10$~MHz, with amplitude of 1~$\mu$m. Then the maximum velocity of the ion is approximately 60~m/s leading to an instantaneous Doppler shift of 125~MHz of the 493 nm beam. The effects of micromotion on Doppler cooling lasers may be less significant for lateral 2D crystals, where lasers can propagate primarily along the trap axis with no micromotion. However, this effect is likely unavoidable for the radial 2D crystals, where a significant portion of   $\vec{k}$ must point along the plane of the crystal, and  where micromotion is always present.

As an ion travels back and forth over a period of  micromotion, laser light is Doppler-shifted by a different amount at each point in the ion's trajectory. This can be seen as the frequency-modulation of the laser light in the ion rest frame, with a modulation depth proportional to the micromotion amplitude.  As the micromotion amplitude is increased, ions may be heated where cooling may be expected, or the laser may be completely off resonance~\cite{DeVoe1989,Berkeland1998}. This is likely the reason that radial 2D crystals so far have been limited to sizes of no more than $\sim$30 ions~\cite{Xie2021,Okada2010,Ivory2020}.

In this paper, we first discuss the impact of micromotion on the interaction of a single ion with Doppler cooling lasers, extending previous models to include the $\Lambda$-system. We use the model to estimate the frequency of a second tone in the Doppler cooling lasers to address ions with differing amounts of micromotion.  We  then demonstrate the ability to Doppler-cool radial 2D crystals of up to 4 shells and over 50 ions using this two-tone scheme--an increase of a full shell or approximately $160 \%$ increase in ion number. The crystals are analysed, extracting individual micromotion amplitudes for all ions. This information is used to find the trap center, as well as to  directly measure the  Mattheiu parameters $q_{x}$ and $q_{y}$.

 \begin{figure}
\includegraphics[width=90mm]{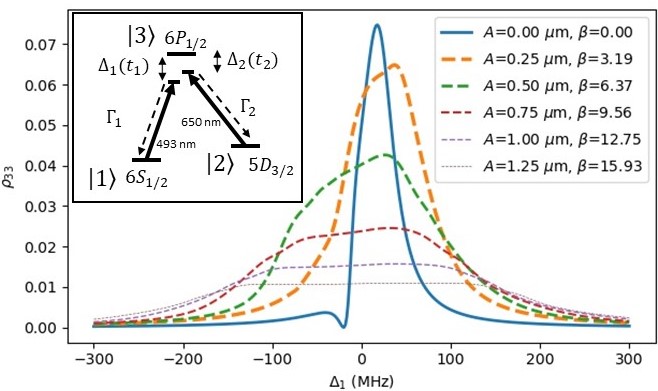}
\caption{\label{fig:Figure_2} Calculated population of the $6P_{1/2}$ state $\rho_{33}$ as a function of detuning $\Delta_{1}$ of the 493~nm laser from the $6S_{1/2}\leftrightarrow 6P_{1/2}$ transition frequency is plotted for six different micromotion amplitudes $A$, indicated in the box in the upper-right, along with the corresponding modulation indices $\beta$. $\Delta_{2}$ is set to -20$\times 2 \pi$~MHz in the lab frame. The scale of the solid line is 1:3, while all of the dashed lines are 1:1.The EIT feature is visible at $\Delta_{2}=-20\times 2 \pi$~MHz in the $A=0.00~\mu$m curve. For non-zero values of $A$, the spectrum is distorted by Doppler shifts due to micromotion, obscuring the  EIT feature. This EIT feature is shifted by different amounts over a period of micromotion, causing ripple features to occur.   The inset in the upper-left shows the relevant energy levels, transition wavelength, decay rates and detunings in $^{138}$Ba$^{+}$.}
\end{figure}

\section{2D crystals, Micromotion, and Laser Cooling}

Consider a single ion in an RF Paul trap displaced at a time-averaged equilibrium position $r_s$ from the center in the crystal plane. Then the equation of motion, assuming that micromotion dominates, is
\begin{equation}
r(t)=r_{s} \left[1+\frac{q}{2} \cos(\Omega t) \right],
\end{equation}
where  $q$ is the Matthieu parameter. The amplitude of excess micromotion is therefore $A=\frac{q r_{s}}{2}$.

For a true Paul trap with hyperbolic electrodes, 
\begin{equation}
q_{r}=q_{z}/2=\frac{8eV}{mR^{2}\Omega^{2}}
\end{equation}
where $e$ is the elementary charge, $V$ is the amplitude of voltage applied to the RF electrodes, $m$ is the mass of the ion, and $R$ the radius of the hyperbolic ring electrode. Therefore, for a fixed  $r_s$, increasing the trap voltage or decreasing $\Omega$ lead to a larger $A$. One can also adjust $A$ by ion species selection or through a change in the size of the trap.
\begin{figure}
\includegraphics[width=80mm]{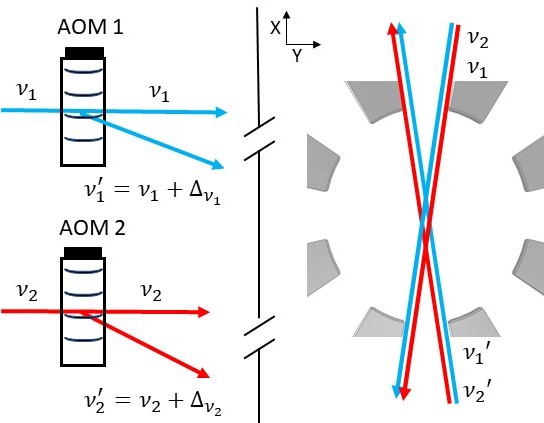}
\caption{\label{fig:Figure_3}Schematic of the two tone laser cooling configuration. The 493~nm and 650~nm beams are split into two distinct frequency tones by AOM's.  The unshifted beams ($\nu_{1}$ and $\nu_{2}$) and blue-shifted beams ($\nu_{1}'$ and $\nu_{2}'$) are combined separately, then counter-propagated through the trap.}
\end{figure}

Since  $A \propto r_{s}$, then for a symmetric crystal $A$ is the same for all ion in each concentric shell. This means there will be two identical regions in each crystal where the projection of micromotion amplitude in the direction of $\vec{k}$ for a particular cooling beam is the same, but with opposite phase. If the crystal is displaced from the origin, the micromotion pattern will no longer be symmetric about the crystal center.
 
When a laser with wavevector $\vec{k}$ is incident on the ion undergoing micromotion of amplitude $A$ about $\vec{r}_{s}$, the frequency of laser light is modulated at RF drive frequency $\Omega$. The frequency deviation due to the micromotion-induced Doppler shift is $\vec{k} \cdot \vec{v}$, where the ion velocity $\vec{v}=\dot{\vec{r}}(t)$. Thus, the frequency modulation index is given by

\begin{equation}
    \beta=\frac{\vec{k}\cdot\vec{v}_{max}}{\Omega} =\frac{\vec{k} \cdot \vec{r}_{s} q}{2},
\end{equation}
where $\vec{v}_{max}$ is the is the maximum velocity of the ion due to the micromotion. When the laser beam is parallel to the direction of the ion displacement $\vec{r_s}$, $\beta=kA$.

To understand the influence of micromotion on laser cooling, one can solve Schrodinger's equations for the $\Lambda$-system and study the behaviour of the excited state population $\rho_{33}$ (see Figure 2 inset). 
For a single two-level ion undergoing micromotion, an analytic solution exists~\cite{DeVoe1989}. However, at higher laser intensities, and in the case of a 3-level atom, the solution must be found numerically. For the two-level system, this can be done by finding the steady state solution for each detuning over a period of micromotion, then time averaging the results~\cite{DeVoe1989}. We extend this model to study a 3-level $\Lambda$-system, which is applicable to the cooling of  ${}^{138}\textrm{Ba}^{+}$ that we trap and laser-cool. In the 3-level case, the numerical approach  must be extended to also time-average over each repump laser detuning $\Delta_{2}(t_2)$ in addition to the main cooling transition detuning $\Delta_{1}(t_1)$, since there is no guarantee that photons from each beam will be absorbed at the same point in the RF period.  Although the spectrum of the laser is different than in the two-level case, we find similar trends.

In Figure 2, the population of the excited state $\ket{3}$ in a 3-level $\Lambda$-system is plotted as a function of lab frame detuning $\Delta_{1}$ of the laser from the atomic resonance frequency for different micromotion amplitudes. We consider a single trapped $^{138}\textrm{Ba}^{+}$ ion Doppler-cooled on the $6 S_{1/2}\leftrightarrow 6P_{1/2}$ transition near 493~nm ($\ket{1}\leftrightarrow\ket{3}$ transition) with a natural linewidth of $\Gamma_{1}=2\pi \times 15$ MHz and repump transition $5 D_{3/2}\leftrightarrow 6P_{1/2}$ near 650~nm ($\ket{2}\leftrightarrow\ket{3}$ transition) with a natural linewidth  of $\Gamma_{2}=2\pi \times 5$ MHz  undergoing micromotion at $\Omega=2\pi \times 10$ MHz, and a saturation parameter for both lasers of $s=10$, similar to our experimental parameters. To avoid the effects of electromagnetically induced transparency (EIT) that can reduce cooling efficiency or cause  ions to go into a dark state, the 650~nm laser is typically red-detuned by 10's of MHz. We numerically calculate the steady-state population $\rho_{33}$ as a function of $\Delta_{1}$ for different amplitudes of micromotion $A$ fixing the lab frame repump laser detuning $\Delta_{2}$ at -20~MHz. Assuming that the natural linewidths $\Gamma_{1}$ and $\Gamma_{2}$ of the $\ket{1}\leftrightarrow\ket{3}$ and $\ket{2}\leftrightarrow\ket{3}$ transitions, respectively, obey $\Gamma_{1}>>\Gamma_{2}$, the sign of the derivative $\frac{\partial\rho_{33}}{\partial \Delta_{1} }$ determines whether heating or cooling will occur, while its magnitude is proportional to the heating or cooling rate.

Distortions to the atomic resonance become pronounced when the Doppler shift due to micromotion is of the order of the power-broadened linewidth of a transition $\Gamma \sqrt{1+s}=\Gamma'$. In terms of the modulation index $\beta$ this condition can be written as $\beta \Omega\thicksim \Gamma'$. As one can see in Figure 2, for $\beta>3$ the absorption line shape starts to deviate significantly from the Lorentzian by becoming lower and broader. The detuning $\Delta_{1}$ that correspond to the largest value $\frac{\partial\rho_{33}}{\partial \Delta_{1} }$, and thus the most efficient cooling, shifts to the red. The  cooling rate decreases quickly, by a factor of $\simeq 10$ (Supplemental Material Figure S.1) for an ion with  \textit{A}=0.25 $\mu$m  as compared to an ion with no micromotion.

As a result, the cooling beam must be farther red-detuned away from resonance for at least some of the ions in order to cool the ensemble enough to promote crystallization. Since the photon scattering rate is proportional to $\rho_{33}$, the off-resonant ions also appear brighter or dimmer in a time-integrated image ~\cite{Xie2021} such as Figure 1.b. Yet ions with micromotion that is perpendicular to $\vec{k}$ do not vary in brightness. When the efficiency of Doppler cooling is lowered due to the distortions of the absorption line shape, the ions may not fully crystallize and appear as rings without fully localizing, or may fail to crystallize at all. When the amplitude of thermal oscillations exceeds the inter-ion spacing, we refer to the crystal as melted.

To address the distortions and cool large crystals, it has been proposed to implement multi-tone laser cooling, or to power broaden the transition and detune the cooling laser~\cite{DeVoe1989}. Power broadening may be limited by available laser power as the size of the crystal grows. For example, in our system, we can achieve a maximum saturation parameter of approximately $s=20$ for a beam waist of 50~$\mu$m. The spectrum in this case is still quite distorted when $\beta=15$, which occurs at $A\simeq1.2~\mu$m. In addition, cooling rates are reduced and the Doppler-limited temperature is increased at higher saturation parameter values. This may lead to the onset of melting of the crystal due to low frequency (soft) normal modes of oscillation, where the amplitude of oscillations can exceed the ion spacing even at temperatures of 10's of mK. We note that over the last 15 years, additional red detuned cooling beams have been used by a rising number of groups to cool ions after large heating events based on empirical evidence of their utility. To the best of our knowledge, this work has not yet been published, and nobody has explored using a second cooling tone to improve cooling of ion crystals where ions experience differing amounts of micromotion. 

We should point out that one may wish to operate in the regime $\Omega >> \Gamma$; in this case the spectrum is divided into micromotion sidebands separated by $n\Omega$~\cite{Berkeland1998}. While this case may be useful, it is experimentally difficult to study since multiple cooling tones would be needed to cool even small 2D crystals, yet it may be easier to manage for larger ion crystals where the laser tones can be evenly spaced. Increasing $\Omega$ may also be impractical for some trap geometries where very high RF voltages would be needed, such as the present experiment. However, this strategy may be suitable for trap geometries with a small spatial size, such as a surface trap~\cite{Wang2020}. For our trap, a factor of two reduction in $\beta$ by increasing $\Omega$ while maintaining reasonable trap strength would require greater than 2000~V amplitude of RF drive, which is typically enough to cause arcing in the RF delivery system. Thus, we do not explore this regime in this paper. We study the two-tone Doppler-cooling of 2D ion crystals with moderate power broadening and demonstrate that larger crystals can be stabilized.

Finally, we note that the steady state picture that we presented in this section is incomplete, since each ion is continuously changing in velocity and hence the amount of Doppler shift. While the steady state solution allows one to compute a laser frequency that results in net cooling over a period of micromotion, in reality, a constant frequency laser tone designed to address ions with a chosen amount of micromotion  only cools optimally for part of each RF phase. 

\section{Experimental System and Methods}

The trap we use is a modification of the original Paul trap, described in~\cite{Ivory2020}. The ring electrode is flattened and divided into eight sectors that allow versatile control of the planar (X-Y plane) trapping potential, while the two endcap electrodes are more closely spaced to yield a stronger transverse confinement. The endcap electrodes are hollowed-out truncated cones that allow sufficient optical access for ion imaging. The planar trap frequencies may be tuned via eight DC voltages on the ring sectors, while the transverse confinement may  be adjusted via DC voltages on the endcap electrodes. We laser cool and trap $\text{Ba}^{+}$ using laser light near 493 nm and 650 nm. RF at $\Omega=2\pi\times10.42$~MHz is delivered to the endcap electrodes via a double-coil helical resonator, resulting in trap voltages in the range 800-1500~V. 

Two tones of Doppler cooling light for cooling beams are counter-propagated through the trap in the $\pm$ X direction as shown in Figure 3, making an angle of $\theta\simeq10 \degree$ with respect to the crystal plane. Both Doppler-cooling lasers at frequencies $\nu_{1}$ (493~nm beam) and $\nu_{2}$ (650~nm beam) are divided into two separate beam paths. One branch of each beam is double-passed through an acousto-optic modulator (AOM), generating a second frequency tone ($\nu_{1}'=\nu_{1}+\Delta_{\nu_{1}}$ and $\nu_{2}'=\nu_{2}+\Delta_{\nu_{2}}$). The AOM's are driven by a signal generators with variable amplitude and frequency (HP8640B), which is then amplified. This allows for the second frequency tone of each beam to be adjusted dynamically while maintaining the beam pointing.

\begin{figure}
\includegraphics[width=80mm]{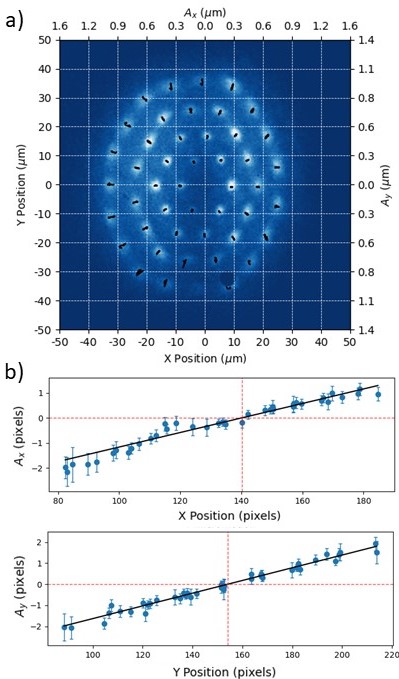}
\caption{\label{fig:Figure_4} Example of data used to determine the location of the trap center and the planar Matthieu parameters $q_x$ and $q_y$ of the trap. (a) A 48-ion crystal, overlaid with the trajectory extracted for each ion over one RF period. There is a dark ion (different isotope of barium) located at the trap center. The dark spot at $(x,y)\simeq(8,-32)$ is a dead spot on the intensifier, which is also present in ion images in Figure 5. The trap secular frequencies are $(\omega_{x},\omega_{y},\omega_{z})=2\pi\times (214,185,788)$ kHz. b) Micromotion amplitude of ions in both the x- and y-direction as a function of the equilibrium position of the ions. The amplitude is allowed to be negative as the  phase changes over the RF null. The straight line is a linear fit to the data where the intercept with the horizontal axis indicates the trap center (red dashed lines).}
\end{figure}
The unshifted beams ($\nu_{1}$ and $\nu_{2}$) and shifted beams ($\nu_{1}'$ and $\nu_{1}'$) are combined separately and counter-propagated through the trap. The beams are focused to a waist of  approximately 50~$\mu$m. Trap frequencies are measured by frequency modulating the RF drive and observing fluorescence dips on a photomultiplier tube (PMT).

The ions are imaged using a 0.28 NA long working distance objective (10x Mitutoyo M Plan Apo), capturing $\sim$2\% of emitted photons. The image is intensified with an intensifier (Photonis Cricket\textsuperscript{TM}) with hi-QE-green photocathode and captured using the Timepix3 camera (Tpx3Cam Amsterdam Scientific Instruments). The intensified camera is single-photon sensitive and provides time-stamping functionality in each pixel with precision of 1.6~ns, allowing for spatial and temporal resolution of images \cite{Fisher2016,Nomerotski2019}. We also use a time-digital-converter (TDC) with  260~ps time resolution, built-in to the camera, to time-stamp pulses that are synchronized with $\Omega$ in order to observe micromotion \cite{Zhukas2021}. The same camera has been used before for studies of ion crystals \cite{Zhukas2021_1}, single photon counting \cite{Ianzano2020,sensors2020} and quantum optics experiments where simultaneous imaging and time-stamping of multiple single photons is required \cite{Yingwen2020,Yingwen2021}.

For the data taken using the Tpx3Cam, we first fold single photon events into a single period $\tau$ of the RF drive (96.0 ns). The position of the ion is tracked over the period by fitting the image of each ion to a rotated elliptical Gaussian for frames that consist of equal increments in time, similarly to previously established methods~\cite{Zhukas2021}. The position over $\tau$ is then fit with a sinusoid according to Eq. 1. The amplitude of micromotion and the equilibrium position are extracted from the fit. Finally, the micromotion amplitude of each ion is plotted against the equilibrium position position and a linear fit is used to extract the slope and intercept with the horizontal axis, yielding the parameter $q_x$ and $q_{y}$ and the location of the trap center. This information is used to determine the scale of $A_{x}$ and $A_{y}$ in the ion images. Figure 4 shows an example of such a measurement. A 48-ion crystal is stabilized in the trap, shown in Figure 4(a), and the micromotion amplitude vs. ion equilibrium position for both X- and Y-axis is shown in Figure 4(b).

$\Delta_{\nu_{1}}$ and $\Delta_{\nu_{2}}$ are empirically determined as follows. We first trap a crystal (3 shells) using only a single, unshifted tone of each cooling beam  at $\nu_{1}$ and $\nu_{2}$, blocking the blue-shifted tones at $\nu_{1}'$ and $\nu_{2}'$ . We measure the amplitude of micromotion for outer ions using the technique described above. We then use the results of these preliminary measurements as a starting point for the estimating $\Delta_{\nu_{1}}$ and $\Delta_{\nu_{2}}$ using the simulations depicted in Figure 2, aiming for a detuning that maximizes $\rho_{33}$.

Next, $\nu_{1}$ and $\nu_{2}$ are adjusted so that the outermost ions appear brightest, indicating where $\rho_{33}$ is maximized. The tones at frequencies $\nu_{1}'$ and $\nu_{2}'$ are then unblocked, and their effect on the crystals are observed while dynamically adjusting $\Delta_{\nu_{1}}$ and $\Delta_{\nu_{2}}$.

\section{Results and Discussion}

 We observe that tuning $\Delta_{\nu_{1}}$ has a pronounced effect on the ability to cool and stabilize large ion crystals, and often causes crystallization or melting to occur as the value is changed.   We observe that increasing $\Delta_{\nu_{1}}$ towards its optimal value appears to make the crystal colder, until it is suddenly heated and melts. This likely occurs as the $\nu_{1}'$ reaches a high enough frequency to heat outer ions in the crystal that have higher micromotion amplitudes. We position $\Delta_{\nu_{1}}$ just below this point. $\Delta_{\nu_{2}}$ is then adjusted around its original setting but observed to affect the crystal to a much lesser extent. We observe that once a crystal with 4 shells has formed, the second tone can then be blocked and the ions remain crystallized for approximately 10 seconds before melting. The crystal will then not reform until the second tone is reintroduced.
 
 We find that setting $\Delta_{\nu_{1}}=101$ MHz and  $\Delta_{\nu_{2}}=66$ MHz allows us to trap and cool larger radial 2D crystals than we were previously capable of with a single-tone Doppler cooling setup. The best results are achieved by slightly compressing the crystal along the axis of propagation of cooling light, in order to reduce the extent of micromotion along that direction. The crystal shown in Figure 4 has 48 ions. By fitting the data of micromotion amplitude vs. position, we find the horizontal intercept (trap center) (X,Y)=(140,153)$\pm$(6,6) pixels and the slope (in-plane Matthieu parameters) ($q_{x}$/2,$q_{y}$/2)= (0.032,0.028)$\pm$(0.001,0.001). 
 
 The largest crystal we are able to stabilize is a 54-ion crystal shown in Figure 5. To our knowledge, this is the largest radial 2D ion crystal ever produced, where all ions in the crystal are efficiently cooled and localized to less than the ions' spacings. The two panels in Figure 5 correspond to the two opposite phases of micromotion with the minimal (a) and maximal (b) crystal spatial extent. 
 
The largest amplitude of micromotion here exceeds 1.5~$\mu$m, which corresponds to a peak-peak oscillation of 6 pixels on the camera image. This peak-peak distance is more than 2 times larger than the diffraction limited spot  size of our system of 1.1~$\mu$m, and is clearly visible in  a time-integrated image (Supplemental Material Figure S.2(a)). In our crystals, $\beta$ ranges from 0 at the center of the crystal to greater than 15 for ions at the top left corner of Figure 5.

\begin{figure}
\includegraphics[width=80mm]{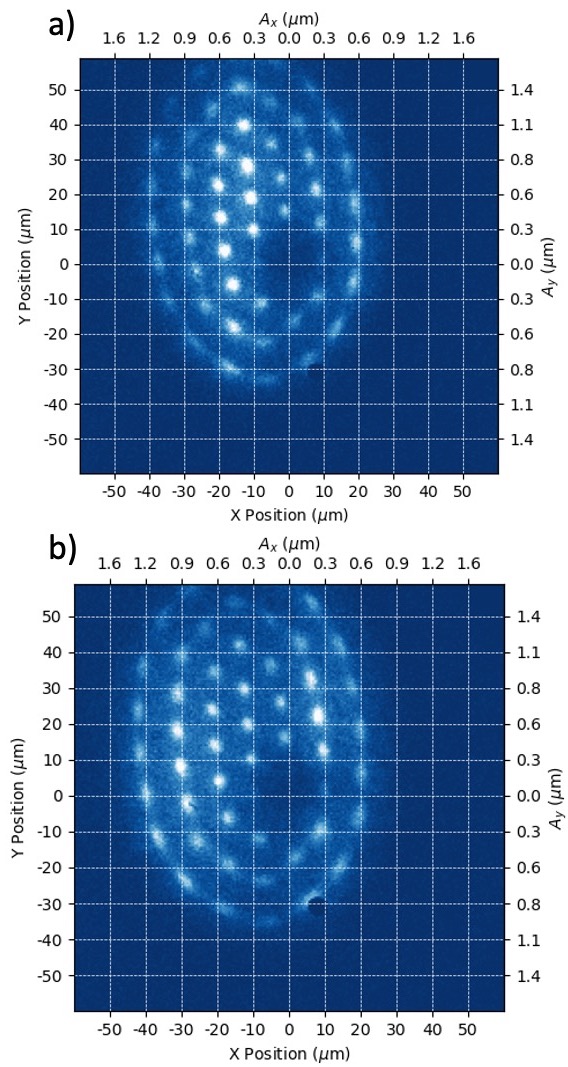}
\caption{\label{fig:Figure_5} A 54-ion radial 2D crystal at different phases of micromotion. There are four dark ions (different isotopes of barium) located near the trap center. The trap secular frequencies are $(\omega_{x},\omega_{y},\omega_{z})=2 \pi \times(203,147,768)$~kHz.  (a) The minimal extent of the crystal that corresponds to the phase of the micromotion when all ions are closest to the trap center. The top-most ion lies outside of the image area for most of the RF period, but can be seen here. (b) the maximal extent of the crystal that corresponds to the phase of the micromotion when all ions are farthest from the trap center. Both images are integrated over time intervals that correspond to 1/36 of the RF period (approximately 2.67 ns).}
\end{figure}

Although the crystals are cooled enough that individual ions can be resolved, thermal motion is clearly visible in the larger crystals. Measuring this temperature may be difficult, since the established methods will likely fail for crystals with significant micromotion. Scanning the cooling laser will not yield useful results, since the response of each ion is different. Methods for thermometry of ion clouds under micromotion have been established~\cite{Sikorsky2017}, but do not extend to the temperature range where crystallization occurs.

We use molecular dynamics (MD) simulations to estimate the temperature of the 54-ion crystal. In our simulations (see Supplemental Material), we compute the trajectories of all ions using the Velocity Verlet algorithm~\cite{Marciante2010}, including micromotion. Damping is introduced to allow the ions to reach their equilibrium positions. Once equilibrium is achieved, the ions are given velocity kicks with random directions and magnitudes from the Boltzmann distribution at a chosen temperature and allowed to reach thermal equilibrium~\cite{Yukai2019}. The trajectories are then analyzed and used to create a time-averaged image where it can be compared  directly to the data. By finding a simulation that yields a crystal with a similar spatial extent of the ion images, we estimate the temperature~\cite{Okada2010}. For the 54-ion crystal, we find that the trajectories are consistent with temperature of $\sim$20~mK, approximately 20x higher than the estimated cooling limit in this power-broadened system.

We also compare this to the spatial extent of outer ions in the crystal with the lowest calculated normal mode frequency~\cite{Landa2012}.  The data collected is folded into a single period of micromotion. In order to separate out spreading due to micromotion from thermal motion, we select data from only 1/16th of the period.   The ion fluorescence profile is the fit to determine the Gaussian rms spread of the ions in the outer shell, taking into account the finite spot size due to our imaging optics~\cite{Blinov2006}. We assume that at equilibrium, the thermal occupation of all modes is equal. Then, according to the spatial extent of the outer ions' motion and the lowest calculated normal mode frequency, the temperature is approximately 20~mK, in good agreement with the result obtained from the simulated images.

The high temperatures are likely the result of a combination of the decrease in cooling rates (Supplemental Material Figure S.1.) due to the distortion of the atomic lineshape, and due to power broadening. Additional tones may help produce larger, or colder crystals using Doppler cooling techniques. However, while additional tones may increase cooling efficiency in some areas of a crystal, they may lead to heating in others. It may also be necessary to control the spatial extent of the laser beams that provide the additional tones to selectively reduce this effect. Finally, a second set of beams at 90\textdegree~ may also further increase cooling capabilities of large crystals.

For larger crystals than studied here, the frequency of the soft modes may drop even further~\cite{Richerme2016} and lead to even lower melting points. This could further limit the extent of power broadening and in order to keep the Doppler temperature below the melting point, and increase the emphasis on multiple cooling tones. Further study is needed to determine the limitations to radial 2D crystal size using Doppler cooling techniques.

The cooling techniques studied here will need to be further improved in order to be useful for quantum information applications. The large amplitudes of thermal motion may contribute to errors in the addressing of qubits by laser light, as individual ions move out of a focused laser beam or other ions move into its path. However, as long as the amplitude of thermal motion is kept smaller than the inter-ion spacing, the crystals should still form. While the amplitude of thermal motion may be large in the plane of the crystal, in the transverse direction, it has been shown that sub-Doppler cooling is possible~\cite{Qiao2021}, and that the transverse motion is well decoupled from in-plane motion~\cite{D'Onofrio2021}. 

In summary, we demonstrate two-tone Doppler cooling of radial 2D crystals of trapped ions. We show that having a second, counterpropagating laser tone helps cool larger radial 2D crystals. We are able to trap, stabilize and efficiently cool crystals with up to 4 shells, and 54 ions. The micromotion of ions in the crystals is analyzed and used to extract the planar Matthieu paramaters $q_{x}$ and $q_{y}$, as well as to locate the trap center.

\section{Acknowledgements}
We would like to thank Megan Ivory for her work on the trap design and fabrication, Peter Svihra for help with the Tpx3Cam  data analysis, and Liudmila Zhukas for help with setting up the camera and the data acquisition. This work was supported by U.S. National Science Foundation award PHY-2011503 and by the BNL LDRD grant 19-30. A.K. acknowledges support under the University of Washington Department of Physics General Excellence Award and J.J.L.and A.G. acknowledge support under the Science Undergraduate Laboratory Internships (SULI) Program by the U.S. Department of Energy.

\bibliography{refs}

\newpage
\setcounter{figure}{0}    

\renewcommand{\figurename}{FIG. S.}

\title{Supplemental Material}


\maketitle
\section{Cooling rates}
In Figure S. 1, the derivatives of the curves seen in Figure 2 in the main text are shown, so as to compare the relative cooling rates at different micromotion amplitudes. Even for ions with just 0.25~$\mu$m of micromotion amplitude, the cooling rate is decreased by a factor of 10 relative to an ion with no micromotion.

\begin{figure}[h]
\includegraphics[width=80mm]{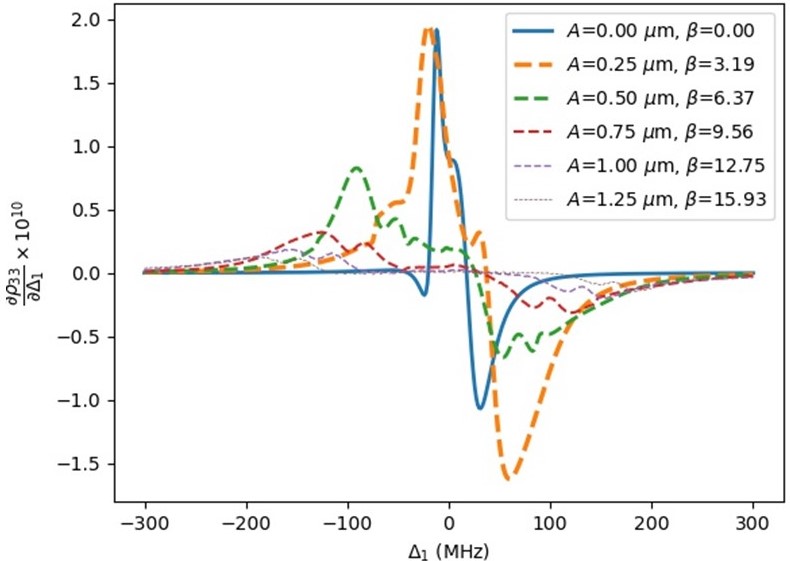}
\caption{Relative cooling rates of ions with increasing amounts of micromotion, as a function of the laser detuning $\Delta_{1}$. The scale of the solid line is 1:10, while are the dashed lines are 1:1.}
\end{figure}

\section{MD Simulations}
Simulations of ion trajectories are used to confirm that various crystal formations are 2D in nature, as well as to estimate the temperature of crystals.

We first simulate the potential in the trapping area by defining the electrode structure and solving for the potential using the Finite Difference Method~\cite{Ivory2020}. Near the trap center, a functional form of the potential is obtained from the numerical values~\cite{Yoshimura2015} to be used in the MD simulations. 

The MD simulations track the trajectories of the ions using the Velocity Verlet algorithm. To ensure this works in the presence of micromotion, small time steps are used (we choose time steps $\tau/20$, where $\tau=2\pi/\Omega$ is the period of the trap RF). Ions are given random initial positions within the trapping area, and velocities are sampled from a Boltzmann distribution at an initial temperature of 100~K. Although the neutral barium atoms are initially at a higher temperature when they are ionized in the trapping area, we find that initializing the  simulation at these higher temperatures does not yield different results, and thus beginning at a lower temperature saves computation time because the equilibrium configuration can be reached in fewer time steps. A velocity-dependent drag force is used to model cooling which is adiabatically turned off over the course of the simulation to find the equilibrium configuration of the ions at zero temperature~\cite{Yukai2019}. To account for displacements of the crystals due to stray electric fields, we introduce a static force to the simulation. 

After the equilibrium configurations are found, the crystals are given velocity kicks. All ions are assigned a random velocity selected from the Boltzmann distribution at a chosen temperature. Then the crystals thermalize over 10,000 RF periods. Finally, an image is generated by tracking the ions position over an array of pixels that represent the physical pixels of the Timepix3 camera. 
We perform the simulations at different temperatures and compare the simulated images with the real one to get an estimate of the temperature of the crystal~\cite{Okada2010}. We find that the extent of ion trajectories is too small at 10~mK, but at 30~mK, the crystal begins to melt. The 20~mK image seems to match our data best, and therefore gives a good estimate of the ion temperature. This is further validated by the comparison with the temperature determined from the thermal oscillations of the ions within the crystal.

\begin{figure}[h]
\includegraphics[width=80mm]{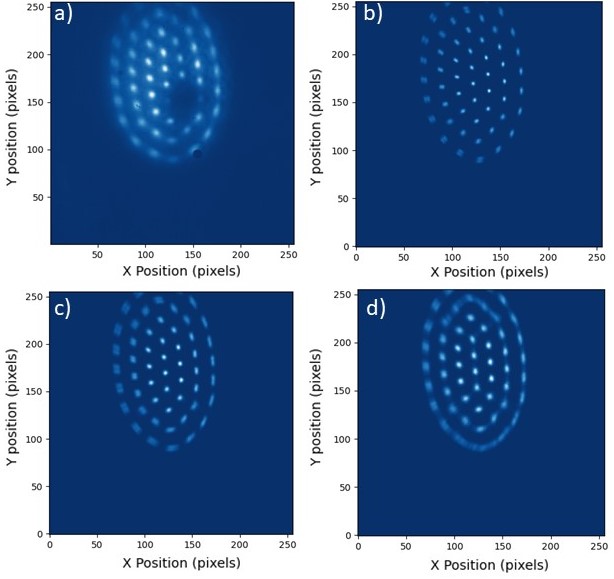}
\caption{Integrated image of 54 ion crystal and simulated images at various temperatures. a) Data taken with the Timpepix3 camera,  the same as used in Figure 5. b) simulated image of the crystal at 10~mK. c) simulated image of the crystal at 20~mK. d) simulated image of the crystal at 30~mK. Here, the onset of radial melting can be seen as the ions in the outer shells begin to delocalize. The point spread function of the imaging optics is not applied to the simulated ion images.}
\end{figure}

\end{document}